\documentclass[aps,prb,twocolumn,groupedaddress,showpacs]{revtex4} 

\usepackage{graphicx}% Include figure files 
\usepackage{dcolumn}% Align table columns on decimal point 
\usepackage{bm}% bold math 
\usepackage{amssymb}

\begin{document}

\title{Implication of the Mott-limit violation in high-$T_c$ cuprates}

\author{Yoichi Ando} 
\email[]{y_ando@sanken.osaka-u.ac.jp}
\affiliation{Institute of Scientific and Industrial Research, 
Osaka University, Ibaraki, Osaka 567-0047, Japan} 

\date{\today}

\begin{abstract}

The Fermi arc is a striking manifestation of the strong-correlation
physics in high-$T_c$ cuprates. In this paper, implications of the
metallic transport in the lightly hole-doped regime of the cuprates,
where the Fermi arcs are found, are examined in conjunction with
competing interpretations of the Fermi arcs in terms of small hole
pockets or a large underlying Fermi surface. It is discussed that the
latter picture provides a more natural understanding of the metallic
transport in view of the Mott-limit argument. Furthermore, it is shown
that a suitable modeling of the Fermi arcs in the framework of the
Boltzmann theory allows us to intuitively understand why the transport
properties are apparently determined by a ``small" carrier density even
when the underlying Fermi surface, and hence $k_F$, is large.

\end{abstract}

\pacs{PACS numbers: 74.25.Fy, 74.72.-h, 74.72.Dn}
%74.25.Fy Transport properties
%74.72.-h Cuprate superconductors (high-Tc and insulating parent compounds)
%74.72.Dn La-based cuprates

\maketitle 

\section{Introduction}

\begin{figure}
\includegraphics[clip,width=5cm]{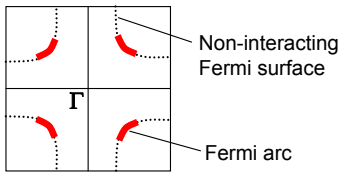} 
\caption{Schematic picture of the Fermi arcs observed in 
lightly-doped LSCO.} 
\end{figure} 

High-$T_c$ superconductivity shows up when carriers are doped to parent
Mott-insulating cuprates where the energy gap opens due to strong
electron correlations, i.e., strong Coulomb repulsions between
electrons. It remains unsolved what exactly happens to the electronic
structure when a small number of holes are introduced into the
Mott-insulating state, although it is intuitively expected that such
holes can move around and give rise to some charge conductivity; indeed,
cuprates show a metallic transport behavior at moderate temperature upon
slight hole doping. \cite{mobility} Recent progress in the
angle-resolved photoemission spectroscopy (ARPES)
experiments\cite{Yoshida1,Yoshida2,ARPES} has elucidated that an unusual
electronic structure called ``Fermi arcs" are formed in the Brillouin
zone at the Fermi energy (see Fig. 1) when the metallic transport
emerges in the lightly hole-doped cuprates. For example, in
La$_{2-x}$Sr$_x$CuO$_4$ (LSCO) with $x$ = 0.03, where a metallic
transport is observed down to $\sim$70 K (see Fig. 2), a well-defined
quasiparticle-like peak is observed in a limited loci in the Brillouin
zone, which defines the Fermi arc;\cite{Yoshida1} intriguingly, these
loci lie on the putative non-interacting Fermi surface that would be
realized in the absence of strong correlations. Since the Fermi arc
starts and ends in the middle of the Brillouin zone, it is not
straightforward to apply such basic notion as the Luttinger sum
rule\cite{Luttinger} that holds in ordinary Fermi liquids. Therefore,
the Fermi arc is a striking manifestation of the new physics that stems
from the strong electron correlations in hole-doped cuprates, and its
understanding is obviously an important step towards understanding the
high- $T_c$ superconductivity.

\begin{figure}
\includegraphics[clip,width=7.5cm]{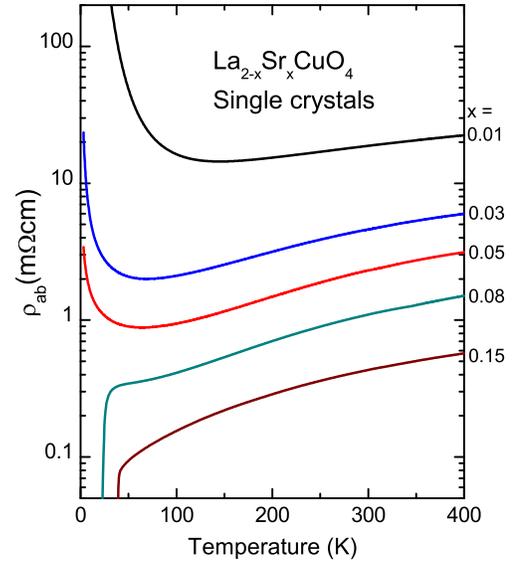} 
\caption{Temperature dependences of $\rho_{ab}$ of a series of 
high-quality LSCO single crystals measured up to 400 K.
Note that a metallic behavior ($d\rho_{ab}/dT > 0$) is observed at 
moderate temperature in all these samples, even for $x$ = 0.01.} 
\end{figure} 

Currently, there are essentially two schools of thoughts to interpret
the Fermi arcs:\cite{ARPES} One is to assume that the hole doping
creates small (closed) hole pockets, but only a half of the pocket can
be seen by ARPES due to a peculiar coherence factor of the
strongly-correlated state. \cite{Chubukov,Chakravarty,Lee} The other is
to consider that there is a large inherent Fermi surface that is screwed
due to the strong correlations, and asserts that only a portion of this
large Fermi surface gains a spectral weight upon slight hole doping and
forms the Fermi arcs. \cite{Norman,Yang,Tohyama,Markiewicz,Stanescu} It
would be very useful if one could elucidate which of the two is likely
to be valid in view of the available transport data. In this article, I
will try to make a case for the large Fermi surface picture based on the
Mott-limit argument. Note, however, that those two pictures both
implicitly assume that the system is uniform, and it will be a different
story if the electrons develop a self-organized inhomogeneity in the
lightly hole-doped cuprates.
\cite{Zaanen,Machida,Muller,anisotropy,Kivelson}

\section{Mott Limit and $k_F \ell$}

An important principle for discussing whether a system should be a metal
or an insulator is the {\it Mott limit} (or Mott-Ioffe-Regel
limit),\cite{Mott} which is based on very simple physics: For the
metallic transport to be realized in a crystalline system, the Bloch
waves of the electrons near the Fermi energy $\varepsilon_F$ must be
well-defined, which dictates that their mean free path $\ell$ must be
longer than the wave length $\lambda_F$ ($= 2\pi/k_F$). This condition
leads to the Mott limit that is expressed as $\ell \gtrsim \lambda_F$,
which is equivalent to $k_F \ell \gtrsim 2\pi$. Since this argument is
quite crude, it is customary to relax the condition a little and
consider the Mott limit to be given by 
\begin{equation} 
k_F \ell \gtrsim 1. 
\end{equation}

\begin{figure}
\includegraphics[clip,width=7cm]{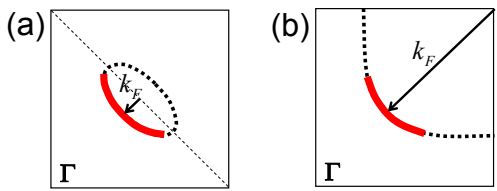} 
\caption{The Fermi wave vector $k_F$ associated with the Fermi arc
in (a) the small hole pocket picture and (b) the large Fermi surface
picture. Here, only a quarter of the 1st Brillouin zone is shown.} 
\end{figure} 

A prominent difference between the small hole pocket picture and the
large Fermi surface picture for the Fermi arc is that the former gives a
small $k_F$, while the latter dictates that $k_F$ must be large, as
shown in Fig. 3. (Note that, for the discussion of the charge transport
by holes, $k_F$ should be measured from the top of the relevant band.)
To examine the metallic transport in LSCO shown in Fig. 2, let us try to
estimate the electron mean free path $\ell$ by using the simple Drude
theory, which gives the following expression for the resistivity:
\begin{equation} 
\rho = \frac{c_0 m^*}{e^2 n_{\rm 2D} \tau}, 
\end{equation}
where $m^*$ is the electron effective mass and $\tau$ is the relaxation
time. Note that in the present case we consider a layered
quasi-two-dimensional system with the layer distance $c_0$, and the
resistivity $\rho$ is measured in the three-dimensional (3D) unit
$\Omega$cm. To calculate $\tau$ from $\rho$ we need the carrier density
$n_{\rm 2D}$, and we assert that it is the carrier density suggested by
the Hall coefficient that should be used for this calculation. This is
because in the lightly hole-doped regime (0.01 $\leq x \leq$ 0.05), as
was demonstrated in Ref. \onlinecite{AndoHall}, the Hall coefficient
$R_H$ is nearly constant at moderate temperature (see Fig. 4) and its
value agrees very well with $c_0/n_{\rm 2D}^{\rm nominal}e$, where
$n_{\rm 2D}^{\rm nominal} = x/a^2$ and $a \simeq$ 3.8 \AA \ is the
in-plane lattice constant. (It is shown in the next section that, based
on the Boltzmann theory, one should use the ``small" $n_{\rm 2D}$ for
this calculation even within the large Fermi surface picture.)

\begin{figure}
\includegraphics[clip,width=8cm]{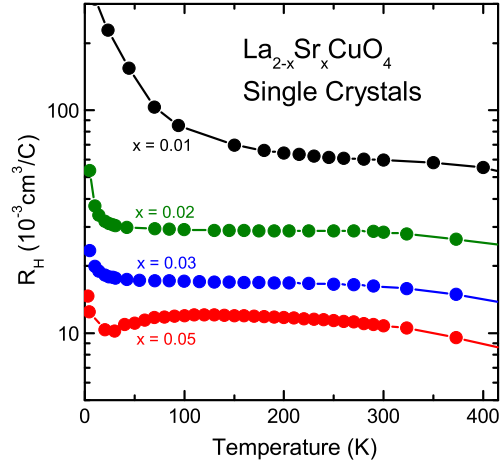} 
\caption{Temperature dependences of $R_H$ of high-quality LSCO single
crystals in the lightly-doped regime ($x = 0.01 - 0.05$) measured up to
400 K. Note that $R_H$ is essentially $T$-independent at moderate
temperature, and its value agrees very well with $c_0/n_{\rm 2D}^{\rm
nominal}e$ (= $a^2c_0/xe$), suggesting that a small carrier density 
$n_{\rm 2D}^{\rm nominal}$ is actually determining the transport
properties.} 
\end{figure} 

In the small hole pocket picture, one may use the conventional formula
$k_F = \sqrt{2\pi n_{\rm 2D}}$ for the two-dimensional (2D) system, which,
together with $\ell = v_F \tau = (\hbar k_F \tau)/m^*$, leads to the simple,
well-known formula
\begin{equation} 
k_F \ell = (h c_0)/(e^2 \rho). 
\end{equation}
In the most extreme case of $x$ = 0.01, $\rho_{ab}$ at 300 K is 20
m$\Omega$cm (see Fig. 2), which gives $k_F \ell \simeq$ 0.1 according to 
Eq. (3). Obviously, the Mott limit Eq. (1) is strongly violated in this 
picture.

On the other hand, in the large Fermi surface picture, one may use the
value $k_F \simeq$ 0.6 \AA$^{-1}$ suggested by the ARPES
data.\cite{Yoshida1} One may also adopt the $v_F$ value from ARPES
experiments which found universal $v_F$ of $\sim$1.5 eV\AA \ that is
nearly independent of doping.\cite{Yoshida2} Together with the estimate
$m^* \simeq 3m_e$ ($m_e$ is the free electron mass) based on optical
conductivity studies,\cite{Padilla} in this picture $\rho_{ab}$(300 K)
of 20 m$\Omega$cm would lead to $\tau \simeq$ 5 fs, which gives $\ell
\simeq$ 10 \AA \ and $k_F \ell \simeq$ 6. This satisfies the Mott limit
and, hence, the metallic transport in the lightly hole-doped regime is
naturally understood. Therefore, the fact that a metallic transport is
realized at moderate temperature even at $x$ = 0.01 points to the
conclusion that the large Fermi surface picture is the more appropriate
one, while the small hole pocket picture bears a paradox.

\section{Fermi Arcs in the Framework of the Boltzmann Theory}

Let us now discuss how the Boltzmann transport theory\cite{Ziman} should
be applied to the system that has Fermi arcs instead of an ordinary
Fermi surface. First, one should remember that in the Boltzmann theory
the charge current $\mathbf{J}$ is calculated as 
\begin{equation}
\mathbf{J} = 2 \int_{{\rm all\ \mathbf{k}\ states}} 
e \mathbf{v_k} g_{\mathbf{k}} d^3 \mathbf{k},
\end{equation}
where $g_{\mathbf{k}} \equiv f_{\mathbf{k}} - f_{\mathbf{k}}^0$ is the
deviation of the distribution function $f_{\mathbf{k}}$ from its
equilibrium $f_{\mathbf{k}}^0$. When $g_{\mathbf{k}}$ is caused by the
action of an electric field $\mathbf{E}$, in the relaxation time
approximation $\mathbf{J}$ can be written as
\begin{eqnarray}
\mathbf{J} &=& \frac{1}{4\pi^3} \int e^2 \tau_{\mathbf{k}} \mathbf{v_k} 
(\mathbf{v_k} \cdot \mathbf{E})
\left(- \frac{\partial f_{\mathbf{k}}^0}{\partial \varepsilon}\right) 
\frac{dS}{\hbar v_{\mathbf{k}} } d\varepsilon \\
&=& \frac{e^2}{4\pi^3 \hbar} \int_{\rm FS} 
\tau_{\mathbf{k}}(\varepsilon_F)v_{\mathbf{k}}
\left(\frac{\mathbf{v}_{\mathbf{k}} }{v_{\mathbf{k}}} \cdot 
\mathbf{E}\right) dS_F.
\end{eqnarray}
It should be noted that $\partial f_{\mathbf{k}}^0 /\partial
\varepsilon$ in Eq. (5) is non-zero only near the Fermi surface, and
this is why the surface integral on the Fermi surface appears in Eq.
(6). The special care we take in the present calculation to accommodate
the Fermi arc picture is that we allow the relaxation time
$\tau_{\mathbf{k}}$ to vary over the Fermi surface.

Assume that $\mathbf{E}$ is along the $x$ axis. For a 2D system,
one obtains from Eq. (6) that
\begin{eqnarray}
J_x &=& \frac{e^2}{4\pi^3 \hbar} \int_{\rm FS} 
\tau_{\mathbf{k}}(\varepsilon_F)(v_x^2 E_x / v) dL_F \\
&=& \frac{e^2}{8\pi^3 \hbar} v_F E_x \int_{\rm FS} 
\tau_{\mathbf{k}}(\varepsilon_F)dL_F,
\end{eqnarray}
where $v = (v_x^2 + v_y^2)^{1/2}$ and $dL_F$ is the line element 
of the Fermi ``surface" in the 2D Brillouin zone.

Adopting the large Fermi surface picture implies that one accepts
the existence of an underlying large Fermi surface, across which the
occupation number of the electronic states changes from one to zero. In
order to incorporate the Fermi arcs in the above framework of the
Boltzmann theory, one may suppose that only on the ``arc" portion of the
underlying large Fermi surface, $\tau_{\mathbf{k}}$ is long enough
($\tau_{\mathbf{k}} = \tau_{\rm arc}$) to allow well-defined
quasiparticles. Furthermore, one may suppose $\tau_{\mathbf{k}}$ = 0
elsewhere on the underlying Fermi surface, which means that the states
outside the Fermi arcs are incoherent. It should be remembered that 
upon performing the energy integral in Eq. (5) to derive Eq. (6), we
relied on the fact that $\partial f_{\mathbf{k}}^0 /\partial
\varepsilon$ is non-zero only near the Fermi surface; the property
$\partial f_{\mathbf{k}}^0 /\partial \varepsilon \neq 0$ still holds at
the underlying Fermi surface even when there are no well-defined
quasiparticles on it (such a surface is called ``Luttinger surface" in
the recent literature\cite{Dzyaloshinskii,Phillips}), so Eq. (8) is 
still valid in this picture.

With the above modeling of the Fermi arcs, one obtains the conductivity
on the Fermi arc as
\begin{eqnarray}
\sigma_x^{\rm arc} &=& \frac{e^2}{8\pi^3 \hbar} v_F \int_{\rm FS} 
\tau_{\mathbf{k}}(\varepsilon_F)dL_F \\
&=& \frac{e^2}{8\pi^3 \hbar} v_F \tau_{\rm arc} \cdot L_{\rm arc},
\end{eqnarray}
where $L_{\rm arc}$ is the total length of the Fermi arcs. If the large
Fermi surface is recovered (i.e., the arcs touch the Brillouin zone
boundary), the conductivity $\sigma_x^{\rm full}$ would be written as
$\sigma_x^{\rm full} = (e^2/8\pi^3 \hbar)v_F \tau_{\rm arc} \cdot L_{\rm
full}$ ($L_{\rm full}$ is the total length of the underlying Fermi
surface), so the conductivity of the Fermi arc system can be expressed
as 
\begin{equation}
\sigma_x^{\rm arc} = p \sigma_x^{\rm full}, 
\end{equation}
where $p \equiv L_{\rm arc}/L_{\rm full}$. Since the ARPES data suggests
that $p$ increases linearly with doping in the lightly-doped
regime,\cite{Yoshida2} Eq. (11) allows us to understand why the ``small"
$n_{\rm 2D}$ determines the conductivity even when the underlying Fermi
surface, and hence $k_F$, is large. Quantitatively, the total length of
the actual Fermi arcs observed in the lightly-doped LSCO is about a
factor of three too long\cite{Yoshida2} to explain the measured
$\rho_{ab}$ with the above simple formalism, which indicates that a more
elaborate modeling of the Fermi arcs, such as a $\mathbf{k}$
dependence of $\tau_{\rm arc}$, is necessary for a more quantitative
description; nevertheless, the present model allows an intuitive
understanding of the transport involving the Fermi arcs in the framework
of the conventional Boltzmann theory.

\begin{figure}
\includegraphics[clip,width=8.5cm]{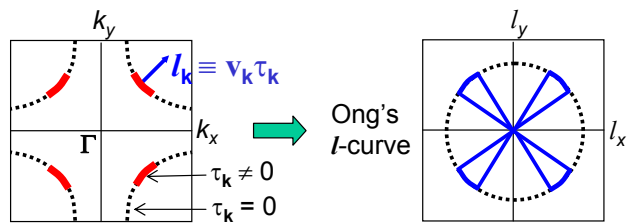} 
\caption{Schematic diagram of the Ong's ``\textbf{\em l}-curve" 
(Ref. \onlinecite{OngHall}) to calculate the Hall conductivity of 
the Fermi arc system in the framework of the Boltzmann theory. 
For simplicity, the underlying 2D Fermi surface is pictured as 
a simple circle.} 
\end{figure}

As for the Hall conductivity $\sigma_{xy}$, the same modeling of the
Fermi arcs leads to a simple result. For 2D metals, $\sigma_{xy}$ is
most conveniently calculated by using the geometrical interpretation
where $\sigma_{xy}$ is given by the ``Stokes area" traced out by the
\textbf{\em l}$_{\mathbf{k}}$ vector which defines the Ong's
``\textbf{\em l}-curve".\cite{OngHall} Since \textbf{\em
l}$_{\mathbf{k}} = \mathbf{v}_{\mathbf{k}} \tau_{\mathbf{k}}$, the
\textbf{\em l}$_{\mathbf{k}}$ vector is non-zero only for those
$\mathbf{k}$'s that are within the Fermi arcs (Fig. 5), and in the
simplest case of a circular underlying Fermi surface and constant
$\tau_{\mathbf{k}}$ on the Fermi arcs, one obtains $\sigma_{xy}^{\rm
arc} = p \sigma_{xy}^{\rm full}$. In this case, the Hall coefficient can
be written as 
\begin{equation} 
R_H^{\rm arc} =
\frac{\sigma_{xy}^{\rm arc}}{(\sigma_{x}^{\rm arc})^2} 
= R_H^{\rm full}/p .
\end{equation} 
This essentially explains why the Fermi arcs can give rise to a Hall
constant that reflects the ``small" $n_{\rm 2D}$. Of course, for a
really quantitative understanding, one needs a more elaborate modeling
of the Fermi arcs.

\section{Insulating Behavior}

It is prudent to mention that the strong localization (or insulating)
behavior is observed at low temperature in the lightly-doped regime.
While at first sight it appears natural for the strong localization to
occur when there are only a small number of carriers, when one
calculates the $k_F \ell$ value at the onset of the localization, one
finds it unusually large. For example, at $x$ = 0.05, $\rho_{ab}(T)$
shows strong localization below 50 K, where $\rho_{ab}$ = 0.9
m$\Omega$cm (see Fig. 2) and the corresponding $k_F \ell$ value is as
large as 13. This is very unusual, because the strong localization is
observed only when $k_F \ell < 1$ in ordinary materials.\cite{Mott}

Such an unusual nature of the localization in the lightly-doped regime
is obviously related to the $\log(1/T)$ insulating behavior that shows
up when the superconductivity is suppressed by high magnetic fields in
underdoped LSCO,\cite{logT,MI} because the $k_F \ell$ value estimated by
using Eq. (3) at the onset of the insulating behavior was as large as 13
for $x$ = 0.15.\cite{MI} It is intriguing that the $\log(1/T)$
insulating behavior in LSCO has been observed for $x \leq$
0.15,\cite{MI} which exactly matches the doping range where the Fermi
arcs are observed by ARPES.\cite{Yoshida2} This coincidence may imply
that the Fermi arcs are inherently insulating at $T$ = 0 in the absence
of superconductivity,\cite{Choy} and the physical origin of the
``insulating" behavior at large $k_F \ell$ values is an important
remaining question.

\section{Summary}

It is discussed that the large Fermi surface picture for the Fermi arcs
allows us to understand the metallic transport observed at moderate
temperature in the lightly hole-doped regime in a natural way; the small
hole pocket picture, on the other hand, leads to a paradoxical
conclusion of strong Mott-limit violation, as long as the system is
considered to be uniform. Also, it is shown that the charge transport
involving the Fermi arcs with large underlying Fermi surface can be
understood in the framework of the conventional Boltzmann theory, by
simply assuming that only on the Fermi arcs the quasiparticle lifetime
is finite; furthermore, the proposed model allows us to intuitively
understand why the ``small" carrier density determines the transport
properties even when $k_F$ is large.

\begin{acknowledgments} 

I would like to acknowledge helpful discussions with A. Fujimori, S. A.
Kivelson, N. Nagaosa, Z. X. Shen, T. Tohyama, I. Tsukada, S. Uchida, and
T. Yoshida. This work was supported by KAKENHI 16340112 and 19674002.

\end{acknowledgments}

\end{document}